\documentclass[preprint]{elsarticle}

\usepackage{amssymb}

\usepackage{lineno}

\usepackage{amsmath}
\usepackage{mathrsfs}
\usepackage{subfigure}
\usepackage{hyperref}

\newcommand{\RR}{{\mathbb R}}

\journal{Elsevier Mechatronics}

\begin{document}

\begin{frontmatter}



\title{Estimation-Based Model Predictive Control for Automatic Crosswind Stabilization of Hybrid Aerial Vehicles\tnoteref{t1}}


\author{Mohamed K. Helwa} \ead{mohamed.helwa@robotics.utias.utoronto.ca}
\author{Adrian Esser} \ead{adrian.esser@mail.utoronto.ca}
\author{Angela P. Schoellig}\ead{schoellig@utias.utoronto.ca}

\address{Dynamic Systems Lab, Institute
for Aerospace Studies, University of Toronto, Canada.} 


\begin{abstract}
In this paper, we study the control design
of an automatic crosswind stabilization system for a novel,
buoyantly-assisted aerial transportation vehicle. This vehicle has several advantages
over other aircraft including the ability to take-off and land in very short
distances and without the need for roads or runways.
Despite these advantages, the large surface
area of the vehicle's wing makes it more susceptible to wind,
which introduces undesirable roll angle motions. The role
of the automatic crosswind stabilization system is to detect the
roll angle deviation, and then use motors at the wingtips to counteract
the wind effect. However, due to the relatively large inertia of the
wing compared to small-size unmanned aerial vehicles and additional input time delays, an automatic crosswind stabilization system
based on traditional control algorithms such as the proportional-integral-derivative (PID) controller results in a response time that is too slow.
Another challenge is the
lack of high-accuracy wind sensors that can be mounted on the vehicle's wing. Therefore, we first design a
wind torque estimator that relies on inertial measurements, and then use feed-forward compensation
to directly correct for the wind torque, resulting in a significantly
faster response. 
We second combine
the proposed estimator with a model predictive controller
(MPC), and compare constrained MPC with
unconstrained MPC for the considered application.
Experimental results show that our proposed estimation-based MPC strategy reduces the response time of the system by around $80$-$90\%$ compared to a standard PID controller,
without the need for adding wind sensors or changing the hardware of the stabilization system.
\end{abstract}


\tnotetext[t1]{This research was partially supported by the NSERC Engage program.}

\end{frontmatter}

\section{Introduction}
\begin{figure}[t]
\begin{center}
\includegraphics[scale=.215, trim = 10mm 10mm 10mm 10mm]{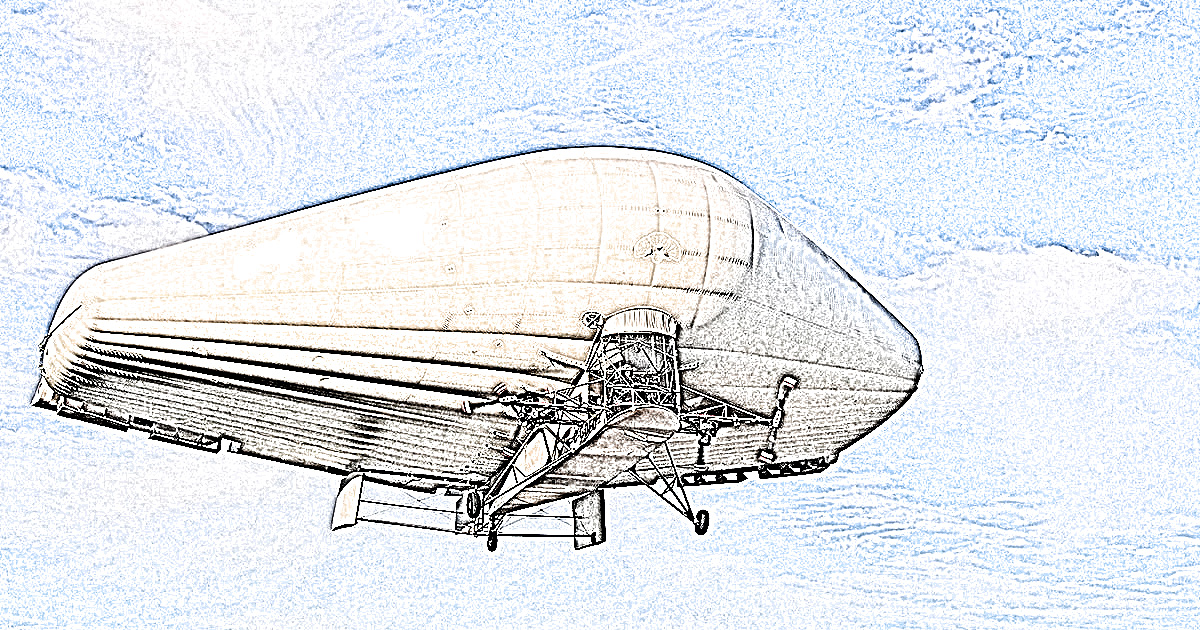} 
\end{center}
\caption{A schematic diagram of the considered hybrid aerial vehicle. In this paper, we design and implement an automatic crosswind stabilization system for the vehicle to counteract the undesirable roll motion caused by wind during take-off, landing, and taxiing.}
\label{fig:solar}
\end{figure}

This paper studies the control design for a novel application, an automatic crosswind stabilization system for a buoyantly-assisted aerial vehicle as shown in Figure \ref{fig:solar}. This novel aerial vehicle design has several advantages over conventional rotary-wing and fixed-wing aircraft, which may be summarized as follows. First, such a vehicle has the ability to take-off and land in very short distances based on the lift generated from the buoyant gas. Second, the vehicle has the ability to carry heavy payloads compared to its own weight; a major part of the lift comes from buoyancy and a small part from aerodynamics. Third, while helicopters depend on fossil fuel, the weight of which can reach 50\% of the helicopter's payload, this novel vehicle design 
is powered by electric engines. 
The vehicle's electric batteries can be recharged in different ways, including solar panels mounted on the wing. 
These advantages, among others, make the novel vehicle design particularly suitable to serve remote areas, most of which have no roads or runways.  

However, despite of its advantages, this novel design results in large surface areas, which make the air vehicle more susceptible to gusts and crosswinds, especially when operating at low speeds during take-off, landing, and taxiing on the ground. In particular, the crosswind causes a roll motion of the wing, which prohibits safe take-off, landing, and taxiing. To solve this problem, an embedded crosswind stabilization system is developed in hardware, which estimates the roll angle of the vehicle's wing using inertial measurement units (IMUs), and then operates electric motors at the vehicle's wingtips to produce a torque counteracting the roll motion caused by the wind. The initial, basic design of the automatic crosswind stabilization system involved the use of a proportional-integral-derivative (PID) controller; however, the response of the system under a well-tuned PID controller is too slow, which prohibits the system from counteracting wind gusts. This sets the stage for the research carried out in this paper.     

The objective of this paper is to design and implement an advanced control strategy for the automatic crosswind stabilization of such novel hybrid aerial vehicles that satisfies the following important properties: 
\begin{enumerate}
\item [\emph{(i)}] fast stabilization of the roll motion of the aerial vehicle under crosswinds and wind gusts, as compared to the basic PID controller; 
\item [\emph{(ii)}] minimal changes of the current hardware of the automatic crosswind stabilization system; 
\item [\emph{(iii)}] implementation on an embedded, microcontroller-based device requiring a computationally efficient  control algorithm. 
\end{enumerate}

In this paper, we provide a computationally feasible, estimation-based model predictive control strategy for this novel application, which reduces the response time of a well-tuned PID strategy by around $80$-$90\%$, without requiring any hardware changes of the automatic crosswind stabilization system.   
 
We first show that a good approach to achieve objective~\emph{(i)} is to measure the torque caused by the wind, and then use a feed-forward compensation to directly compensate for the wind disturbance. However, it is difficult to mount high-accuracy wind sensors on the vehicle's wing and to satisfy objective \emph{(ii)}, we replace wind sensors with a wind torque estimator that relies on inertial measurements. 
We provide two methods for designing the wind torque estimator: the pole-placement technique and Kalman filter \cite{survey_Kalman}.
We then combine the estimator with an improved control algorithm. In particular, we utilize a model predictive control (MPC) strategy \cite{predictive1}-\cite{scenario_mpc}, 
which optimizes the system's future behavior while accounting (directly or indirectly) for system and actuation constraints. 

This paper adds to the wide range of applications of estimators and MPC by utilizing them in this interesting new application. While most of the papers in the literature focus on wind disturbance estimation and rejection for small-size aerial vehicles, as discussed in detail in the next section, we consider a large-size hybrid aerial vehicle and carry out our experimental study on an eleven-meter wing. Moreover, we present and compare, for the considered application, two MPC strategies: a constrained MPC strategy, which requires solving a quadratic program at each sampling instant, and an unconstrained MPC strategy, which reduces to an explicit control law evaluated at each sampling instant. The two MPC strategies are computationally feasible, satisfying objective \emph{(iii)} above. 
Finally, experimental results show the effectiveness of the proposed control strategy.   

This paper is organized as follows. Section \ref{sec:related} summarizes the related work. In Section \ref{sec:model}, we present the dynamic model of the system and discuss how the parameters of the model change depending on the operating conditions. In Section \ref{sec:basic}, we present the basic PID control strategy and illustrate its drawbacks. In Section \ref{sec:proposed}, we present our proposed, estimation-based, predictive control strategy. In Section \ref{sec:exp}, we present experimental results that verify the effectiveness of the proposed strategy and compare the performance of the different control approaches, including unconstrained MPC and constrained MPC. In Section \ref{sec:con}, we conclude the paper. 

\section{Related Work}
\label{sec:related}
In this paper, we study the controller design problem for an automatic crosswind 
stabilization system for a novel hybrid aerial vehicle. Multiple different approaches have been proposed for dealing with this problem in varying contexts. In \cite{turbulence_mitigation_MAV}, a turbulence mitigation system is developed for micro-aerial vehicles (MAVs) using a biomimetically bird-inspired design. Using multiple pressure probes mounted on the wings of the vehicle, the authors of \cite{turbulence_mitigation_MAV} anticipated the properties of the oncoming wind, 
and were able to use this data in a simple feed-forward controller to mitigate the effect of turbulence on the MAV. 
Although this method of anticipating the wind over the vehicles' wings using pressure probes shows promise for small-size aerial vehicles, it does not extend well to large hybrid aerial vehicles. In particular, there are many difficulties associated with mounting reliable pressure probes on our large wing. For instance, these pressure probes do not work well for the low speeds that our vehicle is taxiing at, and thus, the calibration of the sensor must be very accurate and its resolution must be high. As another example, the boundary layer affecting the air flow around the hybrid aerial vehicle is considerably big, which significantly influences the accuracy of these probes when they are mounted on the wing. Thus, we replace the wind sensors in our study with a wind torque estimator.
In~\cite{roll_control}, active wing morphing, together with a simple gain feedback controller, was used for a ten-inch MAV to create a washout effect to aid in damping the wind-induced roll motion. While this method uses a unique control surface to achieve stability rather than anticipating the effect of the disturbances, this technique cannot be efficiently implemented on a large hybrid aerial vehicle due to the prohibitive size and rotational inertia of the airframe.

Kalman filter techniques have been used for wind estimation for other types of
vehicles, as well as for crop protection and building wind load design; see~\cite{multiple_model_filters}-\cite{real_time_muav_imu}.  
In these papers, Kalman filters together with various data mining techniques were used to estimate the wind model, model parameters, and the current state of the wind. Nevertheless, our approach is fundamentally simpler in the sense that we directly estimate the wind torque affecting the roll motion, and rely on only inertial measurements from interoceptive sensors. Also, unlike \cite{multiple_model_filters} and \cite{crosswind_rejection} which contain only simulation results, we provide experimental results on an eleven-meter hybrid aerial vehicle. In~\cite{wind_speed_turbines}, a Kalman filter is used to estimate the wind torque on the rotor of a wind turbine for control purposes. Similar to~\cite{wind_speed_turbines}, we estimate the wind torque on the hybrid airship, but what is unique, compared to \cite{wind_speed_turbines},  is how we use this for roll stabilization and human-in-the-loop control of the vehicle during ground taxiing, landing and take-off. Kalman filtering and model predictive control have been also used for
attitude control and trajectory tracking under aerodynamic disturbances as investigated in \cite{real_time_muav_imu}, \cite{coac_wind_disturbances}-\cite{small_uav_flight_in_wind}. 
These studies are limited though in that they mainly deal with small
systems such as quadrotors or MAVs, and it is not known how these techniques would scale to
larger systems with higher rotational inertias and larger wind disturbances.

While most of the available results in the literature are for in-flight crosswind stabilization of small-size aerial vehicles, our main contribution is to design and implement an efficient automatic crosswind stabilization system for large-size hybrid aerial vehicles during taxiing, and verify its effectiveness experimentally on an eleven-meter hybrid aerial vehicle.

\section{System Model}
\label{sec:model}
We consider the hybrid aerial vehicle in Figure \ref{fig:solar}, and develop an automatic crosswind stabilization system for the scenario when the vehicle is taxiing on the ground. The vehicle considered in this paper is a manned vehicle. During taxiing, the pilot of the vehicle has complete control over the yaw and pitch motions of the aerial vehicle. Through experimental observations, we found that this is not the case for the roll motion under crosswinds. Hence, in this paper, we assume that the pilot is able to keep the pitch angle at zero during taxiing and study the roll motion of the wing. In particular, we design and implement an automatic crosswind stabilization system that detects the roll angle of the wing and then operates motors at the wingtips to counteract the wind torque. 

The objective of this section is to derive a simplified model of the roll motion of the wing when the wing is subject to crosswinds or wind gusts and under the assumption of zero pitch angle. This simplified model will be utilized in our simulations and in Section \ref{sec:proposed} for the design of the estimator and the model predictive controller. 

Although the wing of the considered aerial vehicle is non-rigid, we observed through experiments that the wind torque affecting the wing is not high enough to cause a noticable buckling of the inflated wing. Also, for practical reasons, we seek a simple, finite-dimensional model upon which we can build our estimator and controller. To that end, we assume that the inflated wing is rigid, and utilize Newton's second law for rotational motions to derive an approximate dynamic model of the roll motion of the wing. Let $\theta$ be the roll angle in radians, $\tau_m$ be the total torque of the motors, and $\tau_w$ be the wind torque affecting the roll motion. It is evident that
\begin{equation}
\label{eq:sys_dyn}
J\ddot{\theta}(t)=-K\theta(t)-B\dot{\theta}(t)+\tau_m(t)+\tau_w(t),
\end{equation}
where $t$ is the time index, $J\in \RR$ is the inertia of the wing, $K\in \RR$ is its stiffness (spring constant), and $B\in \RR$ is its damping coefficient. To counteract the wind torque $\tau_w$, two DC motors are mounted on the tips of the wing to produce the counteracting torque $\tau_m$. Both motors are kept at an idle speed and then based on the direction of the wind, one of the motors is sped up to provide a thrust and a counteracting torque. In our proposed setup, each motor is operated in only one direction to avoid the delay associated with reversing the rotational direction of the motor. Hence, we can write
\begin{equation}
\label{eq:motor_torque}
\tau_m(t)=(F_{m,2}(t)-F_{m,1}(t))\frac{d}{2},
\end{equation} 
where $F_{m,1},~F_{m,2}$ are the thrusts of the two motors, and $d$ is the wing span. 
The thrust of the motor $i$ is given by~\cite{Kumar}:
\begin{equation}
\label{eq:motors_thrust}
F_{m,i}(t)=\tilde{K}_{m,i}\omega_{m,i}^2(t),
\end{equation}      
where $\tilde{K}_{m,i}$ is a constant, and $\omega_{m,i}$ is the $i$-th motor speed.
The dynamics of the motors themselves can be represented by \cite{dorf}:
\begin{align}
\label{eq:motors_dyn}
J_{m,i}\dot{\omega}_{m,i}(t)=&~K_{m,i}I_{m,i}(t)-b_{m,i}\omega_{m,i}(t)-\tilde{b}_{m,i}\omega_{m,i}^2(t), \notag \\ 
L_{m,i}\dot{I}_{m,i}(t)=&~V_{m,i}(t) - R_{m,i} I_{m,i}(t)-K_{m,i}\omega_{m,i}(t),
\end{align}
where $I_{m,i}$ is the motor's current, $V_{m,i}$ is its input voltage, $K_{m,i}$ is its torque constant, $b_{m,i}$, $\tilde{b}_{m,i}$ are frictional constants, $R_{m,i}$ is its armature windings resistance, and $L_{m,i}$ is its armature windings inductance. The term $K_{m,i}\omega_{m,i}(t)$ represents the back electromotive force~(emf). 
In practice, the inputs to the above dynamical system are the motor commands sent from a microcontroller to the motor controllers through communication cables, and so it is reasonable to assume that 
\begin{equation}
\label{eq:comm_delay}
V_{m,i}(t)=u_{m,i}(t-T_c),
\end{equation} 
where $u_{m,i}$ is the voltage command sent to to the motor $i$, and $T_c$ is a communication delay. To sum up, the dynamics of the roll motion of the wing with the motors on the wingtips can be represented by equations \eqref{eq:sys_dyn}-\eqref{eq:comm_delay}. Having two states representing the roll angle dynamics \eqref{eq:sys_dyn} and four states representing the two motors, the above dynamics represent a sixth-order nonlinear dynamic system with a time delay. 

The dynamics of the above system can be further simplified as follows. Since the inertia of the two motors is much lower than the inertia of the wing, we neglect the dynamics of the motors in \eqref{eq:motors_dyn}; instead, we assume a longer time delay $T_d$, $T_d>T_c$, between sending the motor commands and the change of the motor torque to the desired value. 
Since \eqref{eq:motor_torque} and  \eqref{eq:motors_thrust} are algebraic equations that can be simply accounted for in the microcontroller code, we assume without loss of generality (w.l.o.g.) that we directly control the motor torque, and hence the dynamics of the system can be represented by:
\begin{equation}
\label{eq:sys_dyn1}
J\ddot{\theta}(t)=-K\theta(t)-B\dot{\theta}(t)+\tau_m(t-T_d)+\tau_w(t),
\end{equation}
where $\tau_m$ is the commanded, total torque of the two motors. Given the physical limits of the motors, the motor torque $\tau_m$ must satisfy
\begin{equation}
\label{eq:saturation}
-\tau_{m,max}\leq \tau_m\leq \tau_{m,max},
\end{equation}            
where $\tau_{m,max}$ is the maximum torque that can be provided by the motors, and $\tau_{m,min}=-\tau_{m,max}$ since the two motors in our setup are identical. Equations \eqref{eq:sys_dyn1} and \eqref{eq:saturation} represent a simplified dynamical model of the system: a linear model with time delay and input constraints. Note that, however, the parameters $J$, $K$, $B$ in \eqref{eq:sys_dyn1} are, in general, time-varying parameters since they depend on the operating conditions of the aircraft such as the ambient temperature and pressure, and the pressure of the wing gas. Since these parameters are changing slowly, we assume that these parameters are fixed during the taxiing, and identify them through simple experimental tests just prior to the start of the vehicle's operation. 


Since the estimator and controller are implemented using a microcontroller, equation \eqref{eq:sys_dyn1} must be discretized. Hence, for a sampling interval $T_s$, we obtain the discrete-time state space model of~\eqref{eq:sys_dyn1},
\begin{eqnarray}
\label{eq:sys_dyn_ss_dis}
x(k+1)&=&A x(k) + B\tau_m(k-k_d) + B\tau_w(k),\notag \\ 
y(k)&=&\left[
1 ~~0
\right]x(k),
\end{eqnarray}  
where $k$ represents the discrete-time index, $x=(\theta,\dot{\theta})$, $A\in \RR^{2\times 2}$, $B\in \RR^2$, and $k_d$ is the number of sample delays.

%
%

\section{Basic Control Design}
\label{sec:basic}
The objective of the automatic crosswind stabilization system is to keep the wing at the horizontal position despite crosswinds and wind gusts; that is, to keep the roll angle of the wing~$y(k)=\theta(k)$ close to zero under the wind torque disturbance~$\tau_w$. At first glance, the problem looks simple. Hence, in this section, we first explore via simulations on MATLAB Simulink the use of a standard PID controller. By analyzing the drawbacks of the standard design, we then propose in the next section an advanced control strategy that significantly improves the control performance.

For our simulations in this paper, we assume $J=6374.5$\,kg.$m^2$, $K=25489$\,N.m/rad, $B=3000$\,N.m.s/rad and $T_d=1$\,s, which are nominal values experimentally identified for our eleven-meter wing prototype with air filling. We also use a wind torque model that relates the wind speed and direction to the resulting wind torque affecting the roll motion, and this model is identified based on computational fluid dynamics (CFD) simulations on the eleven-meter wing. 

\begin{figure}[t]
\begin{center}
\includegraphics[scale=0.5, trim = 25mm 10mm 10mm 10mm]{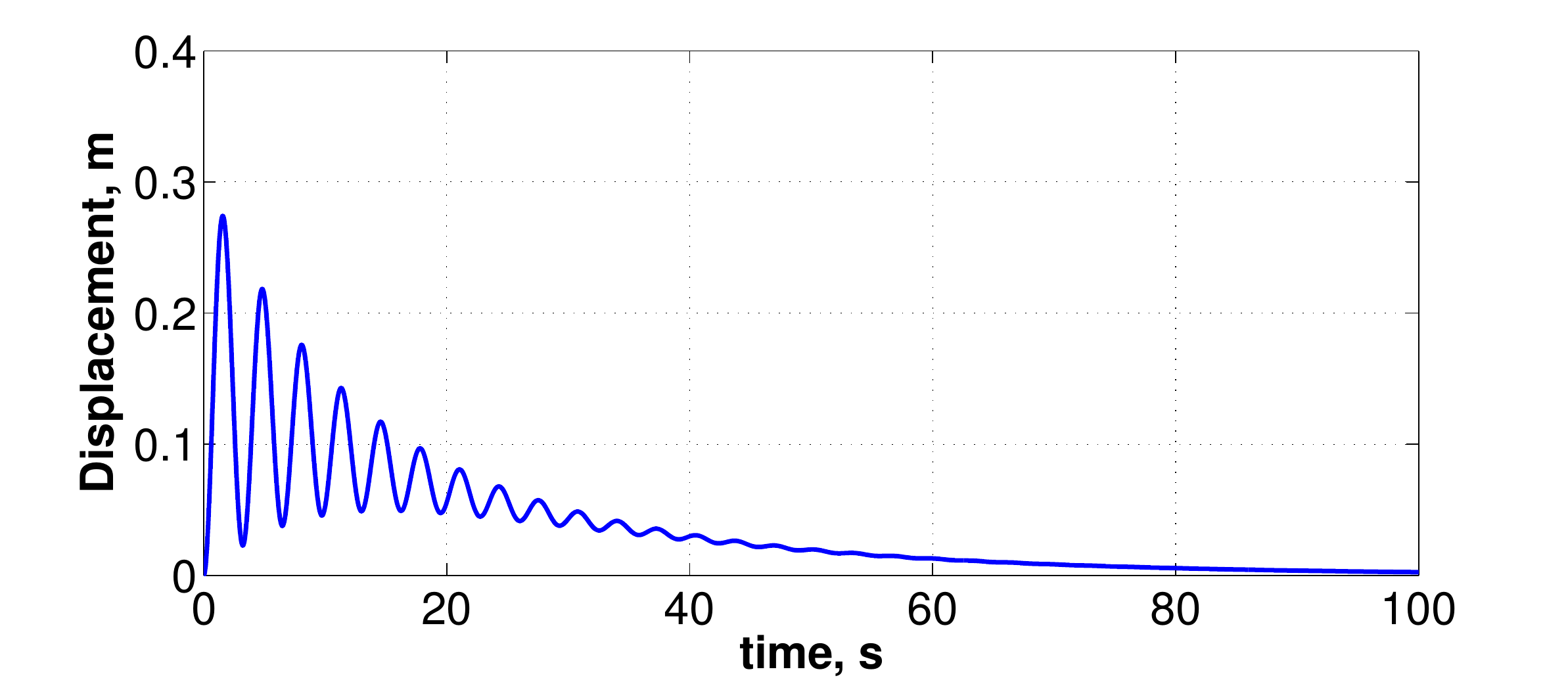} 
\end{center}
\caption{The response of the basic PID controller to steady wind of $7$~$mph$.}
\label{fig_PID1}
\end{figure}

\begin{figure}[t] 
\begin{center}
\subfigure[]
{\includegraphics[trim = 10mm 0mm 10mm 15mm, width = 1\linewidth]{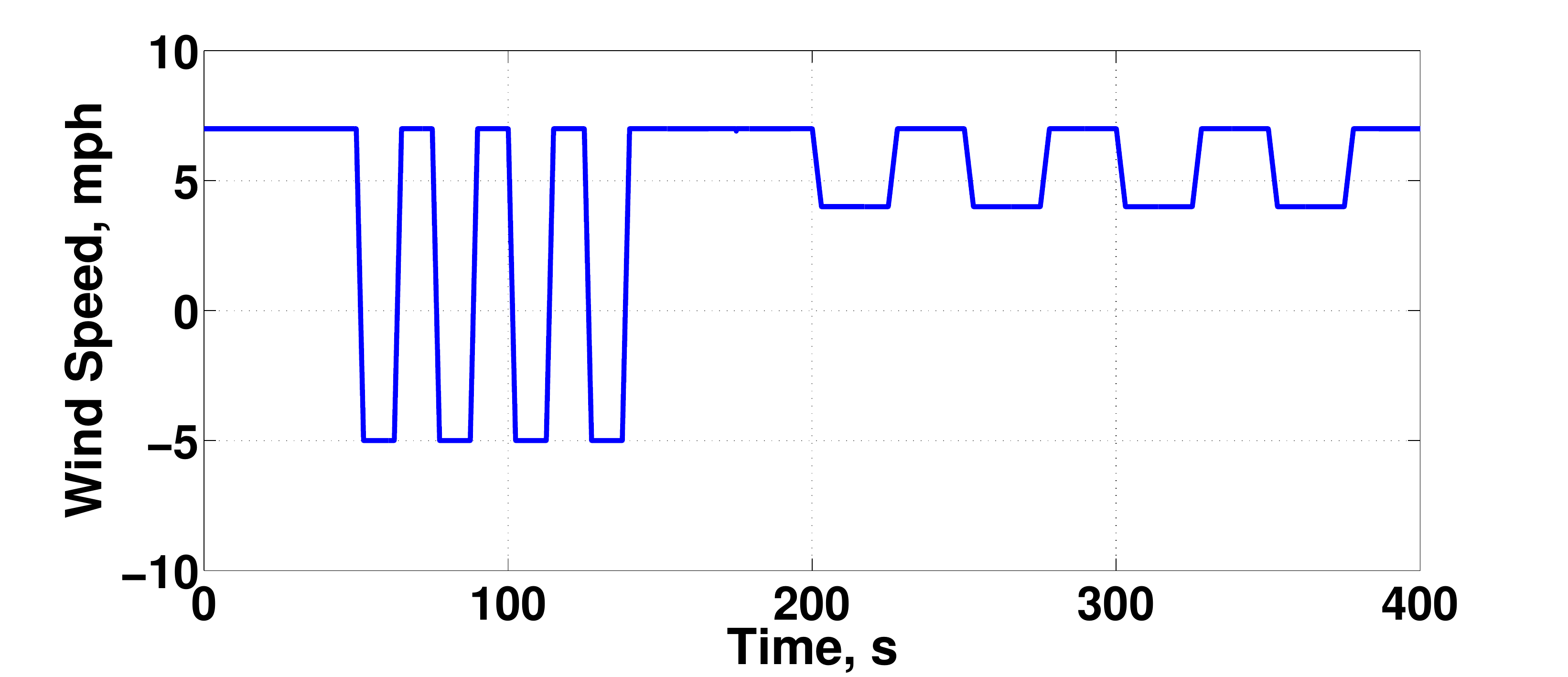}}
\subfigure[]
{\includegraphics[trim = 10mm 0mm 10mm 10mm, width = 1\linewidth]{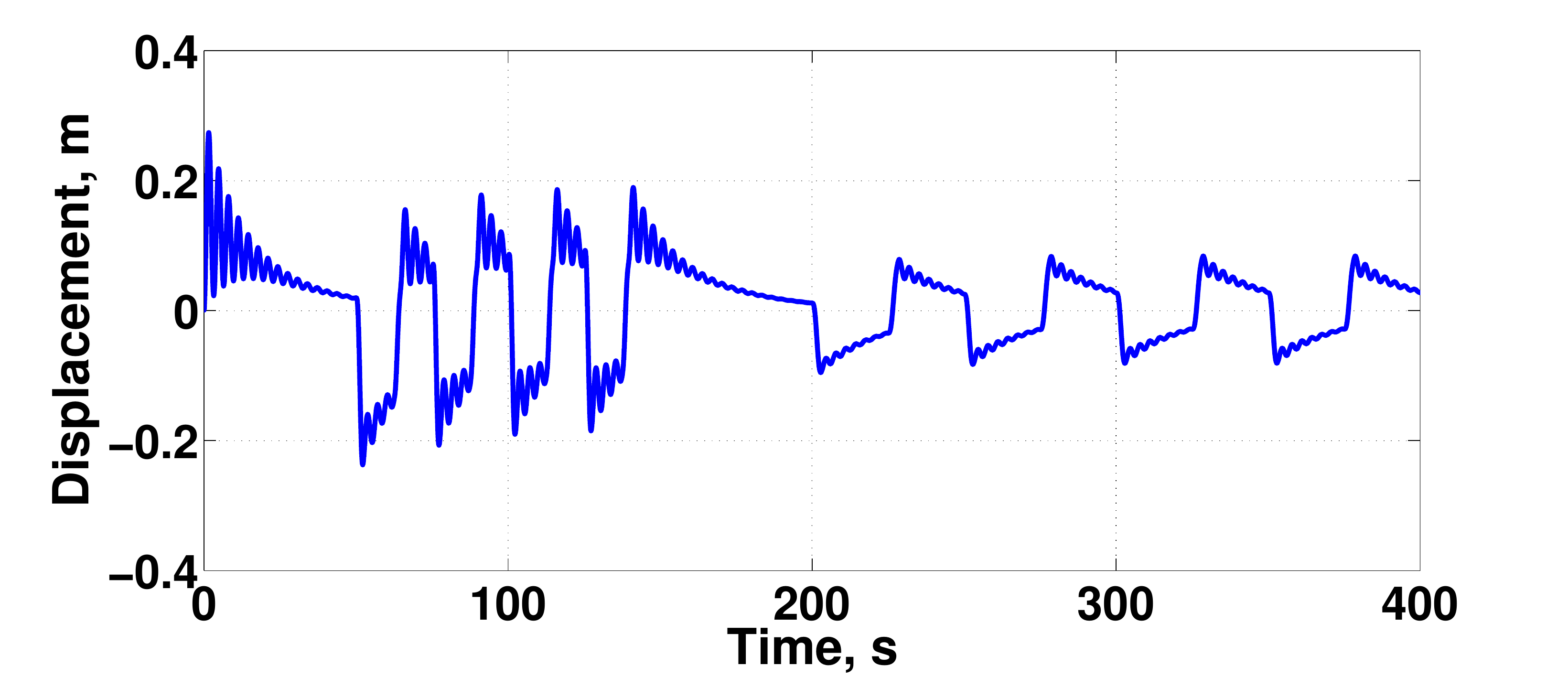}} 
\vspace{-0.2in}
\caption{A wind speed profile containing sudden changes (see (a)), emulating wind gusts, and the corresponding system response under the basic PID controller (see (b)).}
\label{fig_PID_gust}
\end{center}
\end{figure}

For the basic PID control algorithm in our simulations, we select the sampling rate to be $10$ Hz, and the parameters of the PID controller to be $K_p=3200$, $K_I=1200$, and $K_d=700$. The controller gains were selected based on well-known insights into the role of each term of the PID controller and based on extensive trial-and-error. We also add a simple, first-order, low-pass filter before the D-term to avoid noise amplification. Moreover, we use an anti-windup mechanism for the I-term to avoid the windup problem caused by this term if the motor torque is saturated, see \eqref{eq:saturation}. In particular, we saturate the output of the I-term if it hits the actuation limits. Figure \ref{fig_PID1} shows the displacement of the wingtip, given by $\theta~d/2$ where $d$ is the wingspan, for the PID controller under a steady wind of $7$~$mph$.  
Using a standard PID controller, the system is slow and has large oscillations; the settling time is higher than $60~s$. For sudden changes in the wind speed, emulating wind gusts, the controller is not able to return the wing to the horizontal position before the wind changes again. For the wind speed profile in Figure \ref{fig_PID_gust}(a), the corresponding system response is shown in Figure \ref{fig_PID_gust}(b). 
%
%

The inadequate performance of the PID controller is due to the input time delay, as well as the relatively large inertia of the system, causing a slow system response against disturbances. This highlights the need for an advanced control strategy that can sense the wind disturbance and correct for it in a feed-forward fashion, and that can take into account the time delay and the actuation limits.    


\section{Proposed Control Design}
\label{sec:proposed}
In the previous section, we have shown that standard feedback control strategies lead to a slow, oscillatory response against constant crosswinds and wind gusts. When a wind disturbance is applied, the controller first detects the deviation of the angle from the horizontal position, and only then starts changing the motor torque to counteract the wind torque, further delayed by the inherent time delay. The response of the system can be made significantly faster if the wind torque is measured, and directly counteracted by changing the motor torque. This can be achieved by using a wind sensor and a feed-forward compensation technique.  
The crosswind speed can be measured by standard wind sensors such as multi-hole pilot tubes or normal pilot tubes with small vanes \cite{multi_hole_tubes,pilot_tubes2}. 

However, implementing a reliable wind sensor for the large hybrid aerial vehicle has several difficulties. First, 
the wind sensors do not work well for the low speeds that our vehicle is taxiing at, and thus, the calibration of the sensor must be very accurate and its resolution must be high. Second, the boundary layer affecting the air flow around the hybrid aerial vehicle is considerably bigger than the one of a same-size, fixed-wing aerial vehicle. This large effect on the airflow influences the accuracy of the measurements of the wind sensors. Third, the cost of accurate wind sensors is high. Therefore, in our work, we replace the wind sensor with a wind torque estimator. 

\begin{figure}[t]
\begin{center}
\includegraphics[scale=0.6, trim = 8mm 120mm 0mm 120mm]{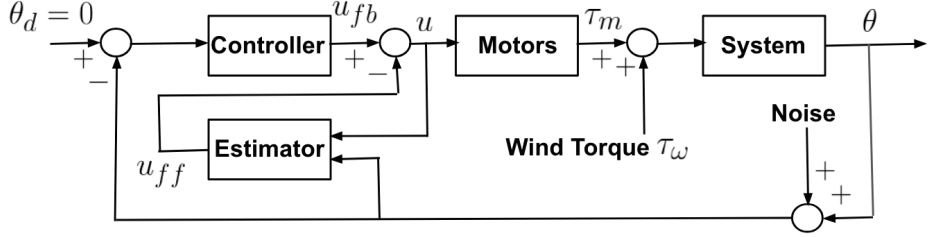}
\end{center}
\caption{A simplified block diagram of the closed-loop control system with a wind estimator and a feed-forward compensation.}
\label{bd_FFS_est1}
\end{figure}

Based on the above discussion, in this paper, we use an estimation-based control strategy as shown in Figure \ref{bd_FFS_est1}. The estimator receives the roll angle measurement and the motor commands, and outputs an estimation of the wind torque that is then subtracted from the controller output to directly compensate for the wind torque disturbance. In the next subsection, we show how to systematically design the wind torque estimator, and then, in Section \ref{sec:mpc}, we present a model predictive controller that takes into account the input time delay and the system's actuation limits. 

\subsection{Estimator Design}
\label{sec:estimator}
Consider the discrete-time state space model \eqref{eq:sys_dyn_ss_dis} of the system, and suppose that the system is affected by an unknown, constant wind torque $\tau_w$. For the short time scale of our stabilization problem, the constant wind assumption in the prediction model is standard and called the `persistence technique' for short-term wind forecasting, see \cite{wind_forecast1}. The objective of this subsection is to design an estimator that can successfully estimate the unknown wind torque from the roll angle measurements and the process model. Since $\tau_w(k+1)=\tau_w(k)$ by assumption, then \eqref{eq:sys_dyn_ss_dis} can be rewritten in the following augmented form: 
\begin{eqnarray}
\label{eq:sys_dyn_ss_dis_aug}
x_{aug}(k+1)&=&A_{aug} x_{aug}(k) + B_{aug}\tau_m(k-k_d),\notag \\ 
y(k)&=&C_{aug}x_{aug}(k), 
\end{eqnarray}  
where $x_{aug}=(\theta,\dot{\theta},\tau_w)$, and the matrices of the augmented system are:
\begin{equation*}
\label{eq:aug_matrices}
A_{aug}=\left[ 
\begin{array}{cc}
A & B  \\
0 & 1   
\end{array} 
\right],~~~B_{aug}= \left[ 
\begin{array}{cc}
B  \\ 0   
\end{array} 
\right],~~~C_{aug}=\left[
1~~0~~0
\right]. 
\end{equation*} 
We have verified that for the nominal system parameters the pair $(A_{aug},C_{aug})$ is observable. Hence, we can design an observer that successfully estimates the states of the augmented system \eqref{eq:sys_dyn_ss_dis_aug} from the output measurements. In particular, we can successfully estimate the unknown value of $\tau_w$ from the output measurements. To that end, let $\hat{x}_{aug}$ be the estimated value of the augmented state in \eqref{eq:sys_dyn_ss_dis_aug}. We use the standard estimator equation \cite{astrom}
\begin{equation}
\label{eq:estimator}
\hat{x}_{aug}(k+1)=A_{aug}\hat{x}_{aug}(k)+B_{aug}\tau_m(k-k_d)+L(y(k)-C_{aug}\hat{x}_{aug}(k)),
\end{equation}    
where $L$ is the estimator gain to be chosen. From \eqref{eq:sys_dyn_ss_dis_aug} and~\eqref{eq:estimator}, the dynamics of the estimation error are:
\begin{equation}
\label{eq:error}
e(k+1)=(A_{aug}-LC_{aug})e(k),
\end{equation}
where $e(k)=x(k)-\hat{x}(k)$. In the following part, we discuss two methods for the proper selection of $L$. 

First, a simple approach is to use pole placement to assign the eigenvalues of $(A_{aug}-LC_{aug})$ to lie anywhere in the open unit disc, and this is always possible since $(A_{aug},C_{aug})$ is observable. This ensures that $e(k)\rightarrow 0$ as $k \rightarrow \infty$, as desired. In our simulations and experiments, we selected the locations of the desired eigenvalues to be $(0.65$, $0.7$, $0.75)$. 

Second, we use the Kalman filter equations to design~$L$~\cite{survey_Kalman,kalman_ss2}. Consider again the augmented system model \eqref{eq:sys_dyn_ss_dis_aug} and suppose that, in practice, noise signals are added to both the state and output equations of \eqref{eq:sys_dyn_ss_dis_aug}, resulting in:
\begin{eqnarray}
\label{eq:sys_dis_aug_kalman}
x_{aug}(k+1)&=& A_{aug} x_{aug}(k) + B_{aug}\tau_m(k-k_d)+v(k),\notag \\ 
y(k)&=&C_{aug}x_{aug}(k)+w(k), 
\end{eqnarray} 
where $v(k)$ and $w(k)$ are zero mean white noise vectors with covariance matrices $Q$ and $R$, respectively. Note that for our problem, $Q$ is a $(3\times3)$ positive semi-definite matrix, while $R$ is a positive scalar. Since the parameters of the wing system are slowly-changing as discussed in Section~\ref{sec:model}, we assume that the matrices $A_{aug}$, $B_{aug}$, $C_{aug}$ are constant during operation, and use a steady-state Kalman filter~\cite{kalman_ss2}-\cite{kalman_ss3}. 
In particular,      
we first solve the algebraic Riccati Equation (ARE) to find the steady-state value of the state covariance matrix~$P$:
\begin{equation}
\label{eq:Riccati}
P= A_{aug}[P-PC_{aug}^T(C_{aug}PC_{aug}^T+R)^{-1}C_{aug}P]A_{aug}^T+Q.
\end{equation}
Then, we use the obtained $P$ matrix to calculate the Kalman filter gain $L_K$ as follows:
\begin{equation}
\label{eq:Riccati}
L_K=A_{aug}PC_{aug}^T(C_{aug}PC_{aug}^T+R)^{-1}.
\end{equation}
The obtained gain $L_K$ can be then used in the observer equation \eqref{eq:estimator} ($L=L_K$). The Kalman filter is optimal in the sense that if the noise in \eqref{eq:sys_dis_aug_kalman} has the distribution as assumed, then the Kalman filter minimizes the mean squared error of the estimated state vector. Even if the noise is not Gaussian, the Kalman filter is the best \emph{linear} estimator for minimizing the mean squared error of the estimated state vector \cite{survey_Kalman,kalman_ss1}. In practice, we typically do not know the covariance matrices $Q$ and $R$. Hence, $Q$ and $R$ are tuned to get a good performance of the estimator. For our aircraft, we select $Q$ to be a diagonal matrix with the following elements on the diagonal: $0.0001$, $0.15$, and $3\cdot10^8$, respectively, and $R=0.01$. 

\begin{figure}[t]
\begin{center}
\includegraphics[scale=0.35, trim = 10mm 10mm 10mm 15mm]{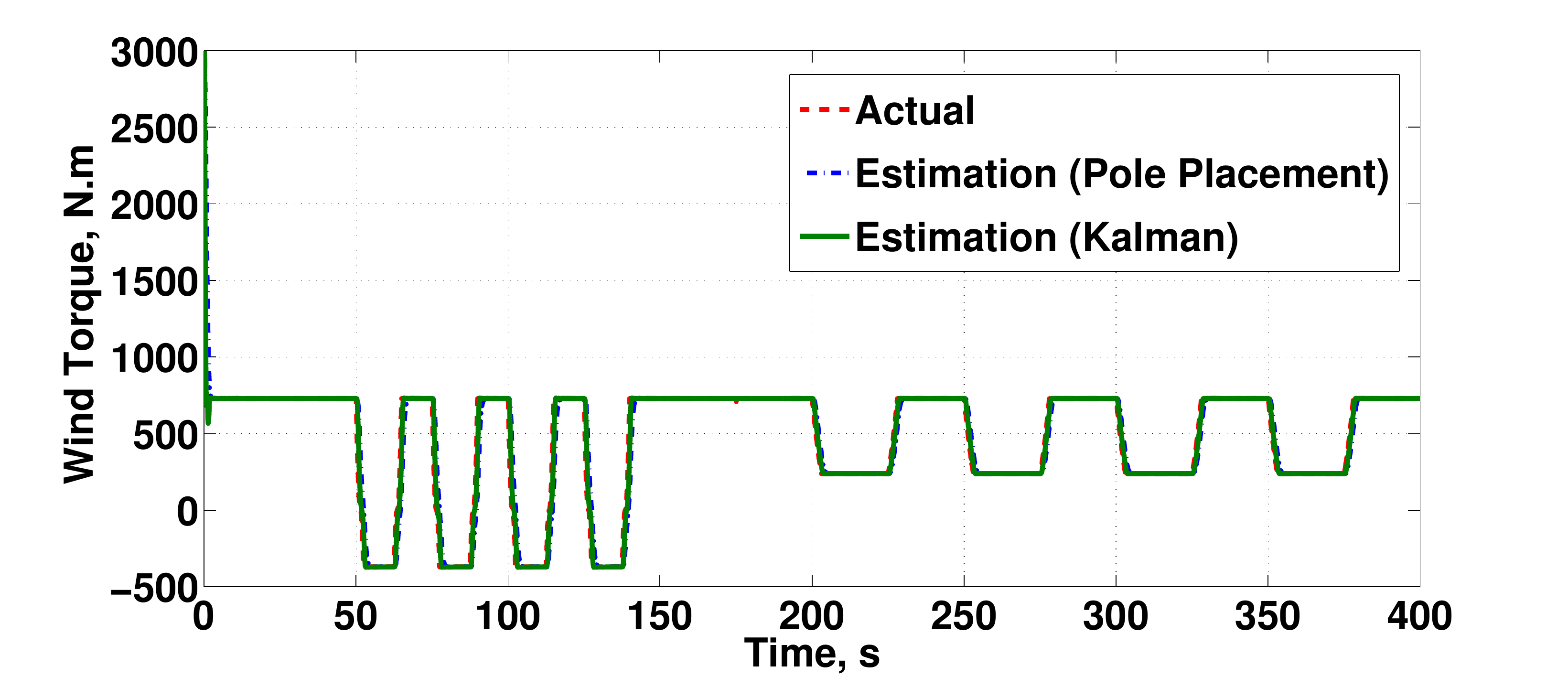} 
\end{center}
\caption{For both design methods of Section \ref{sec:estimator}, the proposed estimator succeeds to estimate the wind torque quickly and accurately under the wind speed profile in Figure~\ref{fig_PID_gust}(a).}
\label{wind_est1}
\end{figure}

Figure \ref{wind_est1} shows the wind torque affecting the wing, corresponding to the wind speed profile in Figure \ref{fig_PID_gust}(a), and the estimated torque for the two proposed methods. In both cases, the estimator succeeds to estimate the wind torque accurately and quickly.

\subsection{Model Predictive Controller}
\label{sec:mpc}
As shown in Figure \ref{bd_FFS_est1}, the estimated wind torque in the previous subsection is utilized to directly compensate for the effect of the wind torque through a standard feed-forward compensation technique. Assuming perfect cancellation of the wind torque using the feed-forward compensation, the discrete-time dynamics of the system become:
\begin{eqnarray}
\label{eq:sys_dyn_ss_dis2}
x(k+1)&=&A x(k) + B\tau_m(k-k_d),\notag \\ 
y(k)&=&\left[
1 ~~0
\right]x(k).
\end{eqnarray} 

The objective of this subsection is to present a model predictive control (MPC) strategy for \eqref{eq:sys_dyn_ss_dis2} to optimize the system's behavior over a prediction horizon of $N_p$ samples.
We start by deriving a prediction model for the system~\eqref{eq:sys_dyn_ss_dis2}. Note that~\eqref{eq:sys_dyn_ss_dis2} has an input time delay, and consequently, one should start the prediction at a time step beyond the value of the time delay since these are the outputs that can be affected by the current and future inputs. To that end, it is straightforward to show:
\begin{equation*}
\label{eq:state_prediction_special}
x(k+k_d)=A^{k_d}x(k)+ \left[
A^{k_d-1}B~~\cdots~~B
\right]
\left[
\begin{array}{c}
\tau_m(k-k_d)  \\
\vdots \\
\tau_m(k-1) 
\end{array}
\right].
\end{equation*}       
At the current sampling instant $k$, we suppose that we have $x(k)$, $\tau_m(k-k_d)$, $\cdots$, $\tau_m(k-1)$ available and, consequently, we can compute $x(k+k_d)$. Let $C=[1~0]$. Then, it can be shown that for all $j=1,\cdots,N_p$, 
\begin{equation}
\begin{split}
\label{eq:output_prediction}
y(k+k_d+j)=CA^{j}x(k+k_d)+CA^{j-1}B\tau_m(k)\\+CA^{j-2}B\tau_m(k+1)+\cdots+CB\tau_m(k+j-1). 
\end{split}
\end{equation}
Collecting these $N_p$ equations in a matrix form, we have
\begin{equation}
\label{eq:predictor}
Y(k)=F(k)+GU(k),
\end{equation} 
where $Y(k)=[y(k+k_d+1),\cdots,y(k+k_d+N_p)]^T\in \RR^{N_p\times 1}$ is the predicted output vector, $U(k)=[\tau_m(k),\cdots,\tau_m(k+N_p-1)]^T\in \RR^{N_p\times 1}$ is the vector of motor control torques to be selected, $F(k)\in \RR^{N_p\times 1}$ and $G\in \RR^{N_p\times N_p}$ are known matrices.

After calculating the matrices $F(k)$ and $G$ at a sampling instant $k$, the constrained MPC problem is
\begin{align}
\label{eq:obj_fun}
& \min_{U(k)} ~Y(k)^TQ_cY(k)+U(k)^TR_cU(k),\notag \\
&subject~to:\notag \\
&U_{min}\leq U(k) \leq U_{max},\notag \\ 
&Y_{min}\leq Y(k)\leq Y_{max},\notag \\
&Y(k)=F(k)+GU(k),
\end{align}  
where $U_{min}\in \RR^{N_p\times 1}$,  $U_{max}\in \RR^{N_p\times 1}$, $Y_{min}\in \RR^{N_p\times 1}$, $Y_{max}\in \RR^{N_p\times 1}$ are constant vectors, $Q_c\in \RR^{N_p\times N_p}$ is a positive semi-definite matrix, and $R_c\in \RR^{N_p\times N_p}$ is a positive definite matrix. The optimization problem \label{eq:obj_fun} is equivalent to:
\begin{align}
\label{eq:obj_fun2}
&\min_{U(k)} ~U(k)^THU(k)+\tilde{F}(k)U(k), \notag \\
&subject~to: \notag \\
&U_{min}\leq U(k) \leq U_{max}, \notag \\
&(Y_{min}-F(k))\leq GU(k)\leq (Y_{max}-F(k)),
\end{align}    
where $H=G^TQ_cG+R_c$ and $\tilde{F}(k)=2F(k)^TQ_cG$. The cost function of \eqref{eq:obj_fun2} is quadratic and the constraints are linear in $U$, and so \eqref{eq:obj_fun2} is a quadratic program that can be efficiently solved online at each sampling instant~$k$~\cite{boyd}. The first element of the obtained $U(k)$ is applied to the system, and the same procedure is repeated at each sampling instant.

The tuning parameters of the MPC are the matrices $Q_c,~R_c$ and the prediction horizon $N_p$. We select $Q_c,~R_c$ as diagonal matrices with positive entries on the diagonal, representing the weight of each output/input in the cost function. The higher the weight, the higher priority is given to minimizing the corresponding variable. We select the last element of $Q_c$ (and hence the terminal cost) higher than the other elements of $Q_c$ to impose $y(k+k_d+N_p)$ to be close to $0$ as desired. As a good initial guess, we select $N_p=N_d-k_d$, where $N_d$ is a reasonably-selected, desired number of samples within which we need the system to reach the desired output. In particular, recalling that the sampling interval $T_s=0.1$\,s, we have $k_d=10$, and we select $N_p=30$.

\begin{figure}[t] 
\begin{center}
\includegraphics[scale=0.35, trim = 10mm 10mm 10mm 15mm]{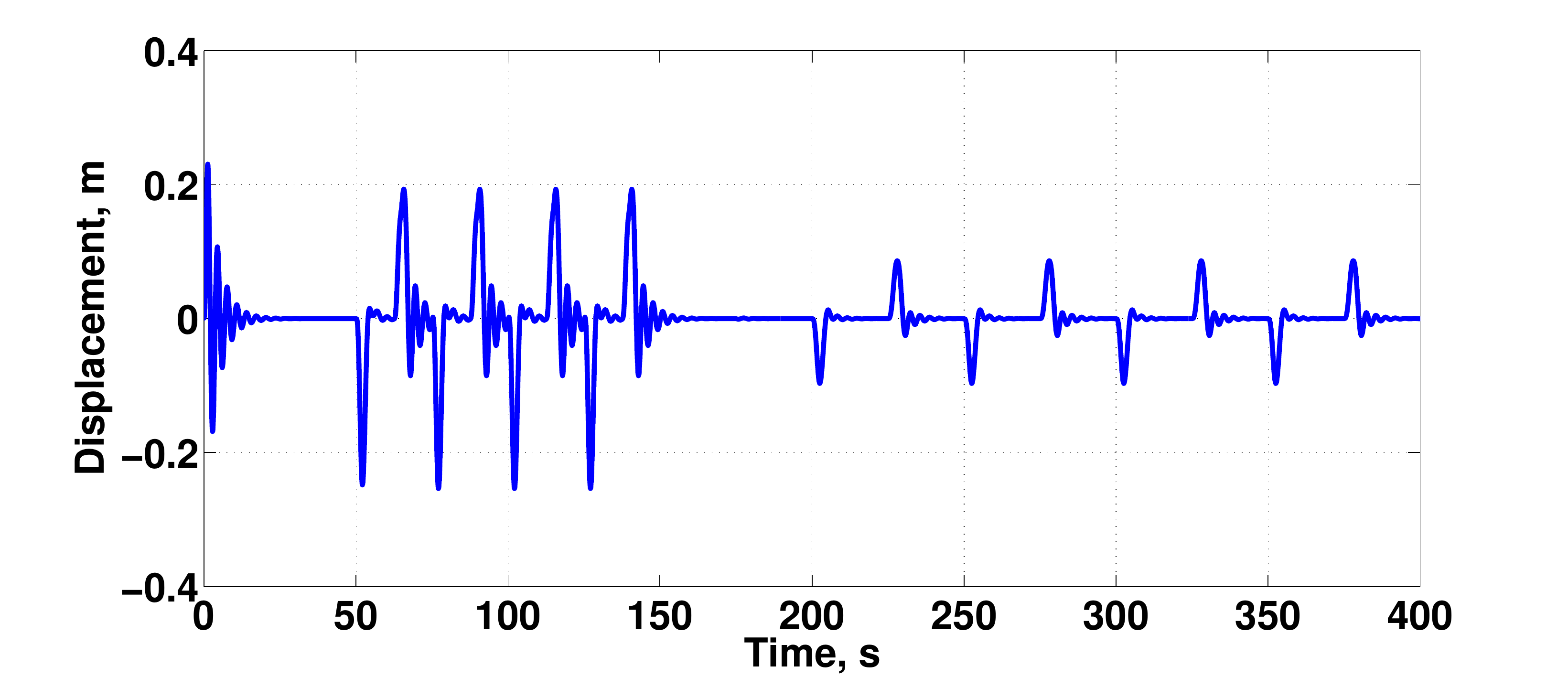} 
\caption{Simulation results for the proposed estimation-based, constrained MPC strategy under the wind speed profile in Figure~\ref{fig_PID_gust}(a). Unlike the PID controller, the proposed strategy quickly stabilizes the wing to the horizontal position every time the wind speed changes. The settling time is around $6-8$ s.}
\label{wind_mpc}
\end{center}
\end{figure}
Figure \ref{wind_mpc} shows the simulated system response under the proposed estimation-based, constrained MPC strategy for the wind speed profile in Figure \ref{fig_PID_gust}(a). Each time the wind speed suddenly changes, the proposed strategy succeeds to stabilize the wing to the horizontal position in about~$6-8~s$, which is around $10$ times faster than the PID control strategy in Figure \ref{fig_PID_gust}(b). The proposed strategy also has lower oscillations than the PID controller.
On the other hand, the constrained MPC strategy requires solving an online optimization problem at each sampling instant, and so it is more computationally demanding than the PID controller. 

Hence, we next present an estimation-based, unconstrained MPC strategy that still achieves significant improvements over the basic PID design, and does not require solving online optimization problems. This approach is useful if only limited online computational resources are available in the automatic crosswind stabilization system. To that end, we ignore the hard inequality constraints on $Y(k)$ and $U(k)$ in the constrained MPC problem, and indirectly account for these constraints by proper selection of the cost matrices $Q_c$ and $R_c$. By ignoring the constraints of \eqref{eq:obj_fun2}, it is evident that the solution of the optimization problem~\eqref{eq:obj_fun2} is a closed-form control law given by:      
\begin{equation}
\label{eq:u_opt}
U_{opt}(k)=-H^{-1}G^TQ_c^TF(k).
\end{equation}  
Note that $H$ is always invertible since $R_c$ is positive definite and $Q_c$ is positive semi-definite by selection. The matrix $H$ is typically of high dimension; however, it is only a function of the system matrices and the cost matrices, and consequently, its inverse can be pre-computed offline to further reduce the online computational burden. Similar to the constrained MPC approach, only the first element of $U_{opt}(k)$ is applied to the system, and then the computation of $U_{opt}(k)$ in \eqref{eq:u_opt} is repeated at each sampling instant.



\section{Experiments}
\label{sec:exp}
In this section, we briefly describe the setup of the experiments, and then present experimental results that verify the effectiveness of the proposed estimation-based MPC approach for the considered application. Finally, we provide a detailed discussion of the results.
\subsection{Experimental Setup and Results}
\label{sec:exp_res}
\begin{figure}[t] 
\begin{center}
\subfigure[]
{\includegraphics[trim = 0mm 0mm 0mm 15mm, width = 0.6\linewidth]{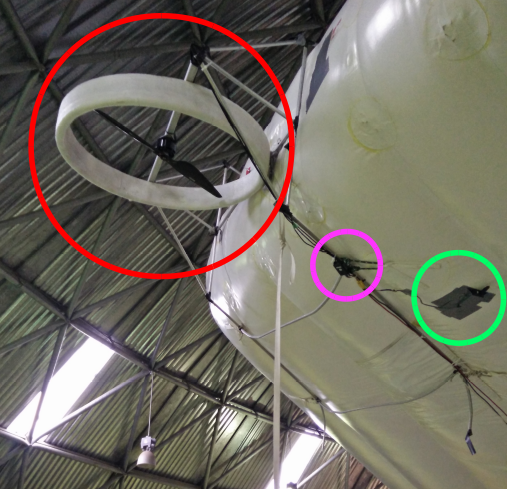}}
\subfigure[]
{\includegraphics[trim = 0mm 0mm 0mm 15mm, width = 0.315\linewidth]{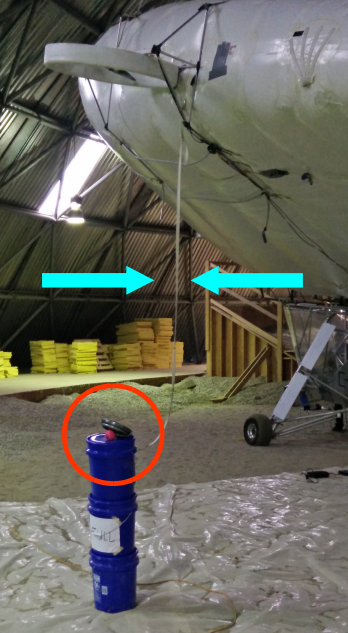}} 
\vspace{-0.13in}
\caption{The figure illustrates the experimental setup: (a) the wingtip components used in our experiments including the motor and propeller (red circle), the motor speed controller (magenta circle), and the inertial measurement unit (IMU) (green circle), and (b) the weight (red circle) used for simulating wind torques in our experiments.}
\label{fig:exp_setup}
\end{center}
\end{figure}

We present experimental results on an $11$-meter prototype of a hybrid aerial vehicle. 
The experiments were carried out at the MarsDome of the University of Toronto Institute for Aerospace Studies (UTIAS). 
In these experiments, the constant wind speeds were simulated with weights attached with a rope at the vehicle's wingtips, while wind gusts were simulated with large, sudden changes in the weight disturbance that can be achieved by putting on and off weights during the experiment. Two inertial measurement units (IMUs) are placed close to the vehicle's wingtips to measure the roll angle of the wing, and we emphasize that these are the only available sensors in our experiments, see Figure \ref{fig:exp_setup}.


\begin{figure}[t] 
\begin{center}
\subfigure[]
{\includegraphics[trim = 10mm 5mm 10mm 15mm, width = 0.7\linewidth]{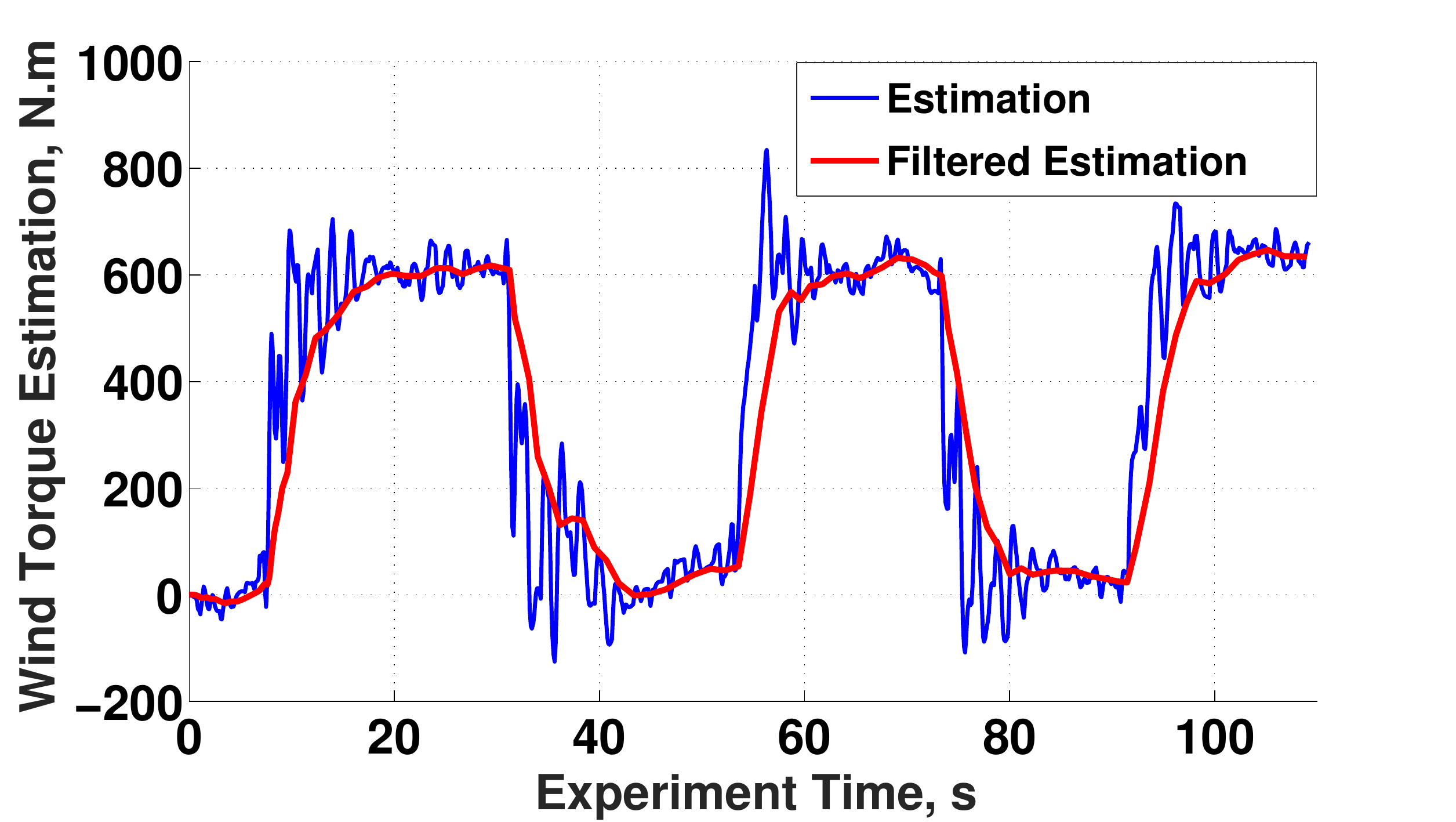}}
\subfigure[]
{\includegraphics[trim = 10mm 5mm 10mm 10mm, width = 0.7\linewidth]{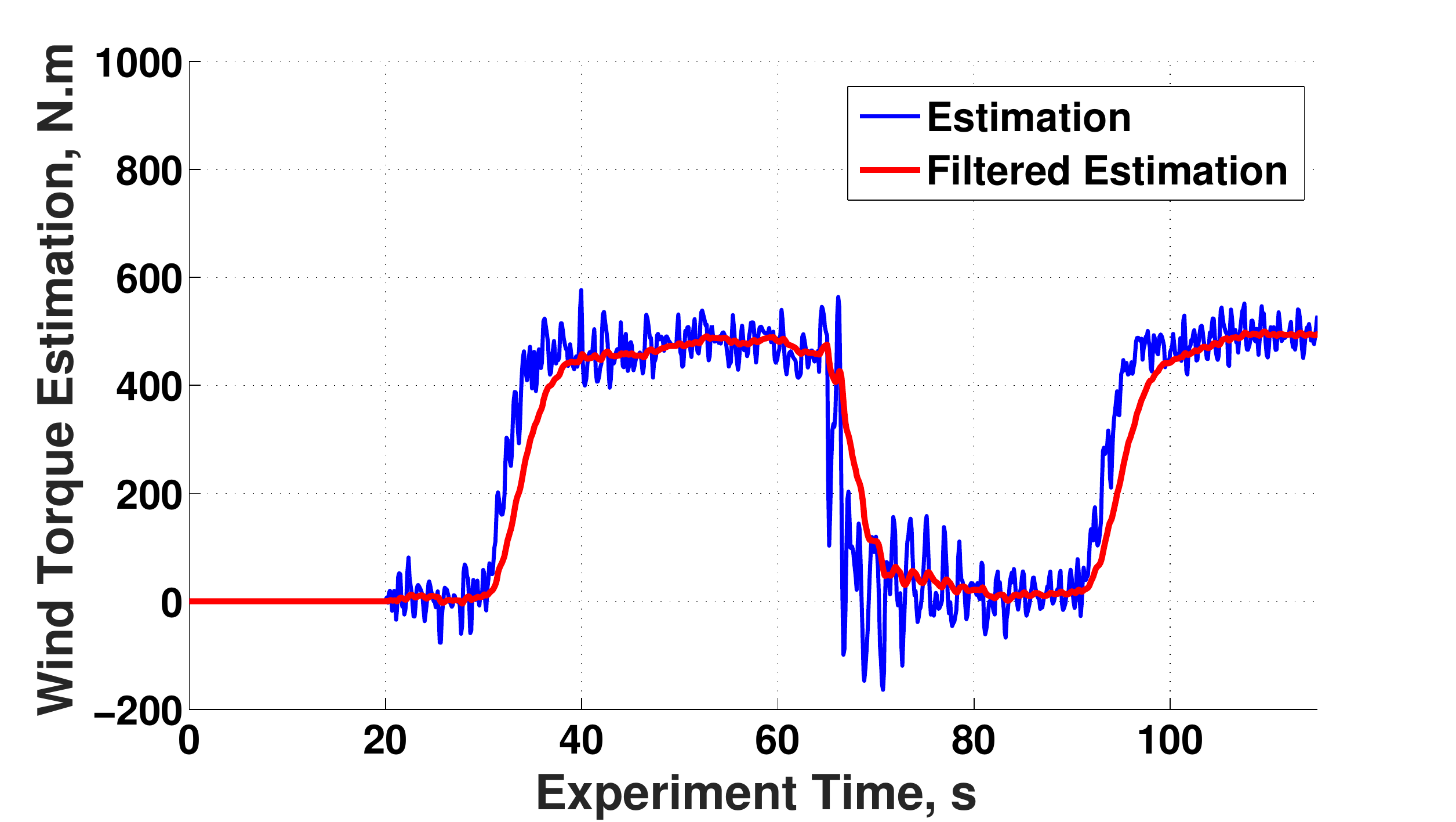}} 
\vspace{-0.1in}
\caption{The wind torque estimation (blue) and its filtered value (red) when putting on and off a constant weight at the left wingtip: (a) $25$~lb weight applied (torque changes of approximately $600$~N.m) with the estimator gain designed based on the pole placement technique; (b) $20$~lb weight applied (torque changes of approximately $500$~N.m) using a Kalman filter.}
\label{fig_wind_est_exp}
\end{center}
\end{figure}

We start with verifying the effectiveness of our proposed wind torque estimator. To that end, we carried our experiments in which we put on and off known weights at the left wingtip, and then detected the corresponding torques affecting the wing using the proposed estimator. Figure \ref{fig_wind_est_exp}(a) shows the wind torque estimation for putting on and off $25$ lb of weight, which corresponds to torque changes of approximately $600$~N.m. For this experiment, the estimator gain is designed based on the pole placement technique discussed in Section \ref{sec:estimator}. One can see that the estimator successfully detects the fast changes in the weights, and it estimates the disturbance torque value almost accurately. However, one can see that the estimated torque is very noisy due to the noise in the measured angle, and consequently, we always use a filtered value of the estimated torque (red curve). 
Figure \ref{fig_wind_est_exp}(b) shows the wind torque estimation for putting on and off $20$~lb of weight, causing disturbance torque changes of approximately $500$~N.m, where for this experiment a Kalman filter is applied with the same parameters used in our simulations in Section~\ref{sec:estimator}. The estimator successfully detects the fast changes in the weight and the filtered, estimated torque is close to accurate.

\begin{figure}[t]
\begin{center}
\includegraphics[scale=0.25, trim = 10mm 10mm 10mm 20mm]{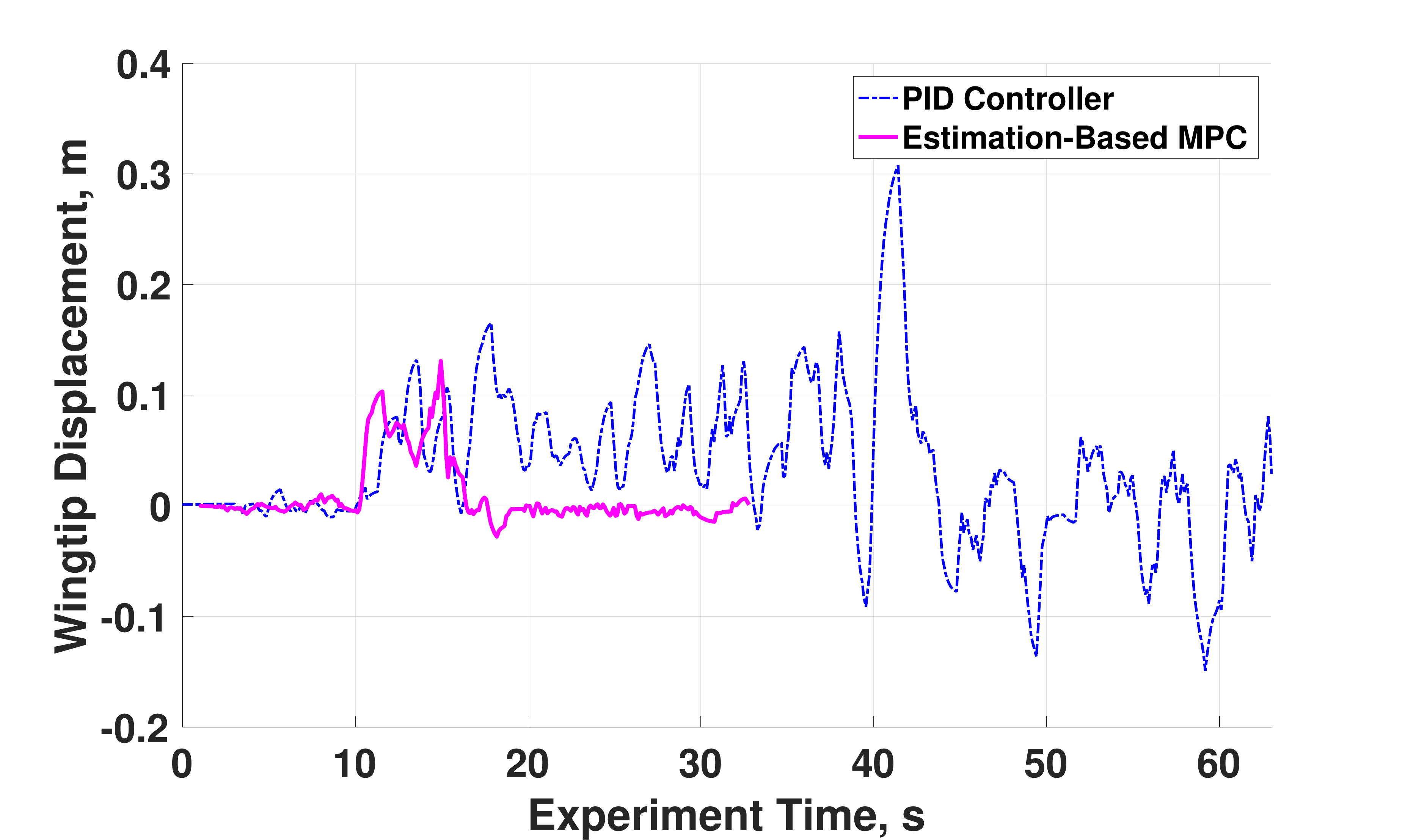} 
\end{center}
\caption{The wingtip displacement for a weight of $15$ lb attached at the left wingtip at time $t=10$ s. The proposed estimation-based, constrained MPC strategy has significantly less oscillations, compared to the PID controller, and it brings the wing displacement back to zero in around $6$ s.}
\label{fig_constant_disp}
\end{figure}

After verifying the effectiveness of the proposed wind torque estimator, we next test the proposed estimation-based, constrained MPC strategy, and compare it to a basic PID control strategy 
\[
CO(k)=K_Pe(k)+K_II(k)+K_D\Delta e(k),
\]
where $CO(k)$ is the output of the PID controller at sampling instant $k$, $e(k)=\theta_d(k)-\theta_f(k)=-\theta_f(k)$ is the error in the roll angle at $k$, $\theta_f(k)$ is a filtered value of the roll angle measurement $\theta(k)$, and $I(k)$ is the integral term given by
\[
I(k)=I(k-1)+T_se(k),
\]  
where $T_s$ is the sampling interval, and we saturate $I(k)$ beyond the actuation limits to avoid the windup problem of the I-term. For the D-term, we select $\Delta e(k)$ to be the average of the Euler approximation of the error derivative over three samples to reduce the effect of noise. That is, we use $\Delta e(k)=(de(k)+de(k-1)+de(k-2))/3$, where $de(k)=(e(k)-e(k-1))/T_s$. Figure \ref{fig_constant_disp} shows the displacement of the wingtip under the two control strategies and for a constant weight of 15 lb added at time $t=10$~s. Based on crosswind experiments on the $11$-meter wing, the torque caused by the $15$~lb is equivalent to the effect of $8$~km/hr crosswind speed. The blue trajectory shows the system response under the basic PID controller, with the parameters $K_P=3200$, $K_I=1200$ and $K_D=700$, which are the same parameters used in the simulation in~Figure~\ref{fig_PID1}. From Figure~\ref{fig_constant_disp}, one can see that the response with the PID controller is very oscillatory and, for around $60$~s, the wing does not settle back to zero displacement, i.e., the PID control strategy fails to stabilize the wing to the horizontal position. On the other hand, the magenta trajectory is the result of our proposed control strategy, and one can see that it has much less oscillations. It settles in around 6~s and has a significantly lower peak value. A demo video is available at \url{http://tiny.cc/EMPC-HybridAircraft}.

\begin{figure}[t]
\begin{center}
\includegraphics[scale=0.25, trim = 10mm 10mm 10mm 20mm]{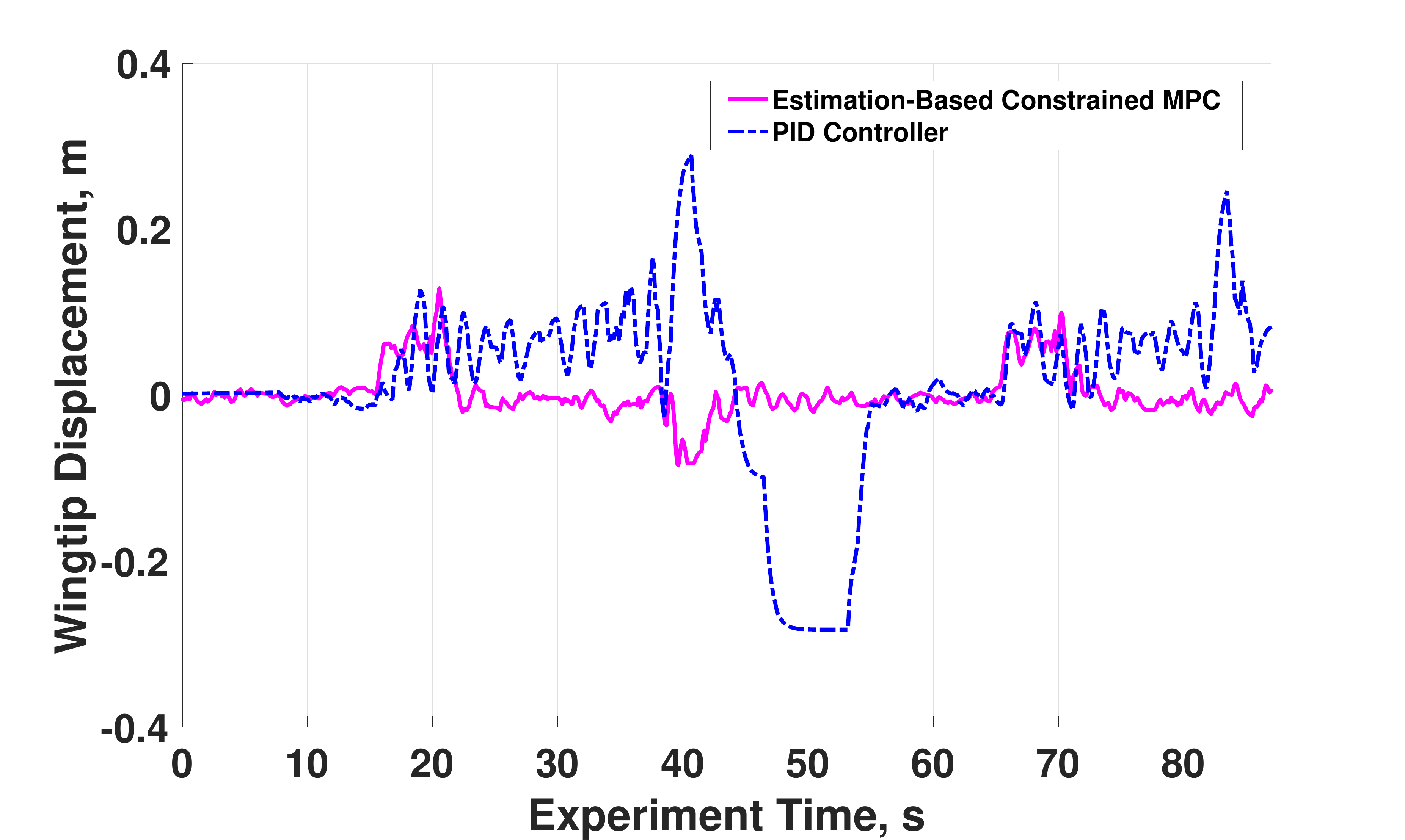} 
\end{center}
\caption{The wingtip displacement for a square disturbance resulting from putting on and off a weight of $15$~lb at the left wingtip every $25$\,s starting from time $t=15$\,s. The proposed estimation-based, constrained MPC strategy has significantly less oscillations compared to the PID controller, and it stabilizes the wing displacement to zero quickly and before the weight changes again.}
\label{fig_square_disp}
\end{figure}

\begin{figure}[t]
\begin{center}
\includegraphics[scale=0.325, trim = 10mm 10mm 10mm 5mm]{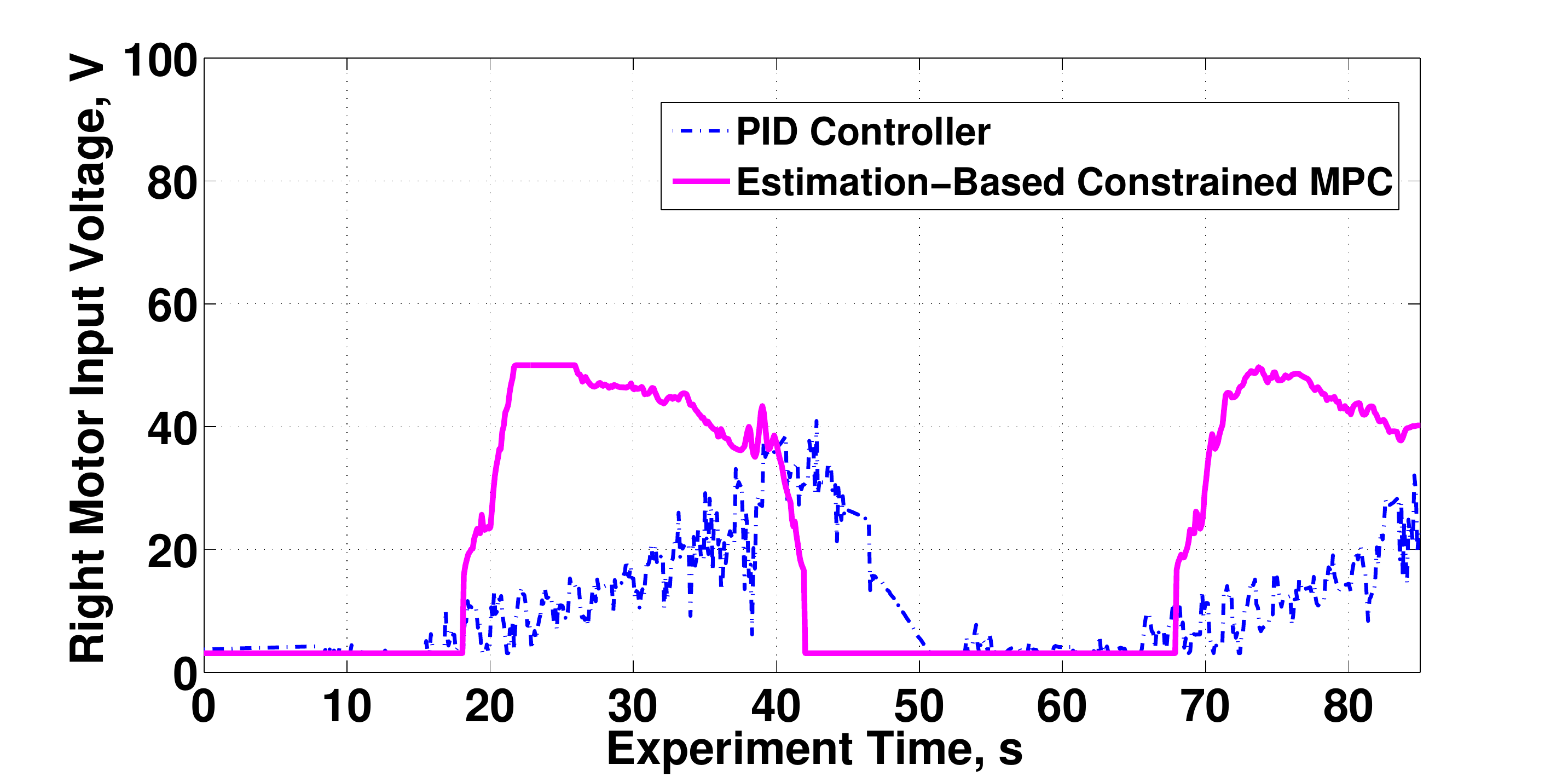} 
\end{center}
\caption{The right motor input voltage commands for a square disturbance resulting from putting on and off the $15$ lb weight at the left wingtip every $25$\,s starting from time $t=15$\,s. The dashed blue lines are for the basic PID controller, while the solid magenta lines are for our proposed estimation-based, constrained MPC strategy. The right motor must speed up when the weight is put on at the left wingtip. The motor commands in our proposed strategy react faster and are less oscillatory.}
\label{fig_square_motors_cmd}
\end{figure}

We second test the ability of the counter gust system to respond fast to wind gusts, which are mainly characterized by fast changes in the wind speed. To simulate this in the MarsDome, we put on and off the $15$ lb weight approximately every 25 s, and the counter gust system must handle this change in weight. Figures \ref{fig_square_disp} and \ref{fig_square_motors_cmd} show the wingtip displacement and the motor commands, respectively, for the same two controllers as in Figure \ref{fig_constant_disp}, particularly the basic PID controller (blue) and the proposed estimation-based, constrained MPC (magenta). 
One can see from Figure~\ref{fig_square_disp} that the proposed estimation-based strategy was able to stabilize the wing to zero displacement quickly (in around $6$\,s) every time the weight changes, and the response has significantly less oscillations and lower peak values. Also, from Figure~\ref{fig_square_motors_cmd}, the motor commands in the proposed estimation-based MPC strategy reach suitable values faster every time the weight changes, and they oscillate less resulting in a smoother response.

\begin{figure}[t]
\begin{center}
\includegraphics[scale=0.25, trim = 13mm 10mm 10mm 20mm]{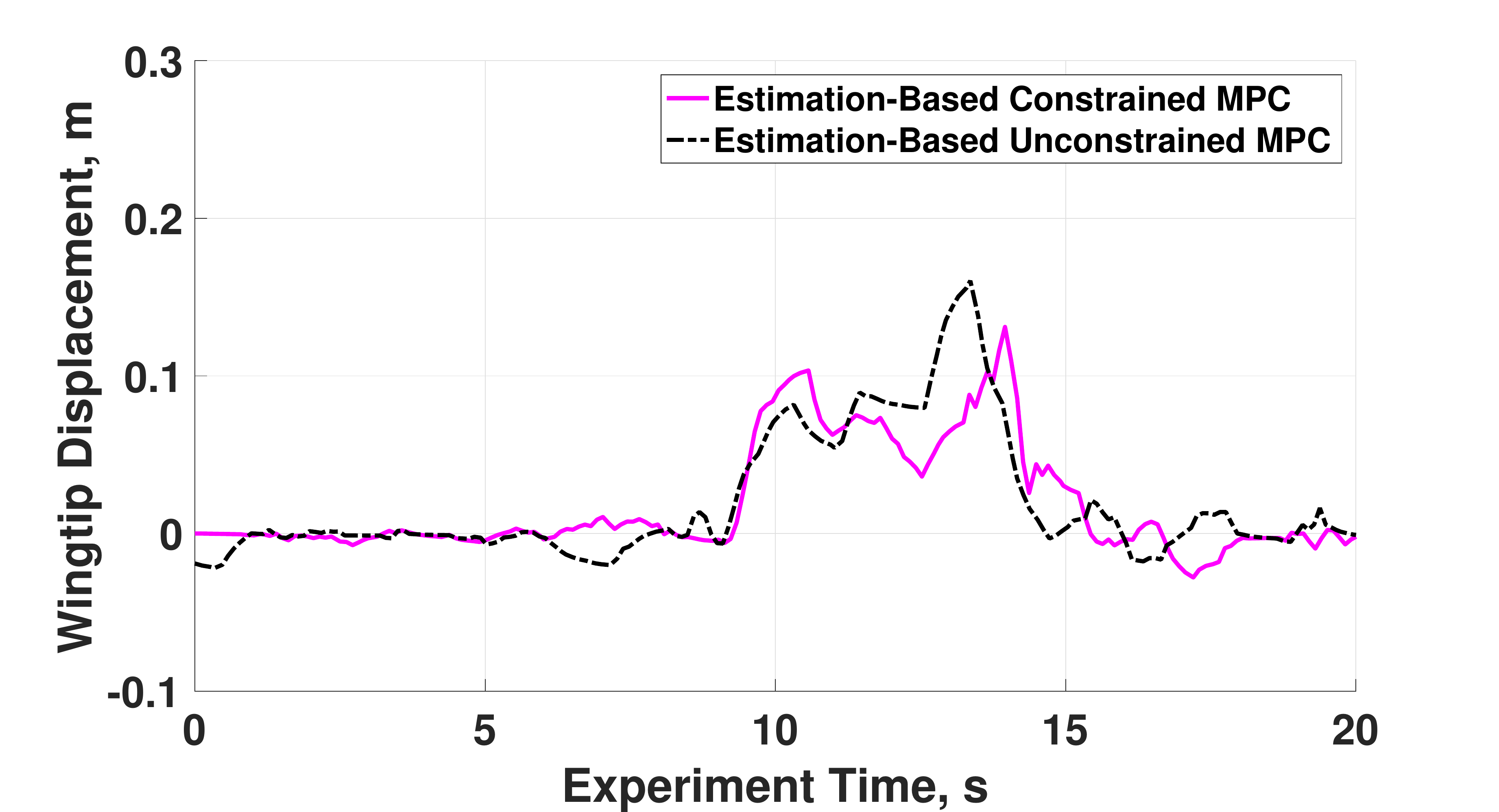} 
\end{center}
\caption{The wingtip displacement for a weight of $15$~lb attached at the left wingtip at time $t=9$ s. The proposed estimation-based, unconstrained MPC strategy achieves a fast response; it settles the system in around 5 s. It reaches slightly higher peak value than the estimation-based, constrained MPC. However, it requires significantly less computational power compared to the constrained MPC.}
\label{constrainedMPC_versus_unconstrained}
\end{figure}

\begin{figure}[t]
\begin{center}
\includegraphics[scale=0.25, trim = 10mm 10mm 10mm 10mm]{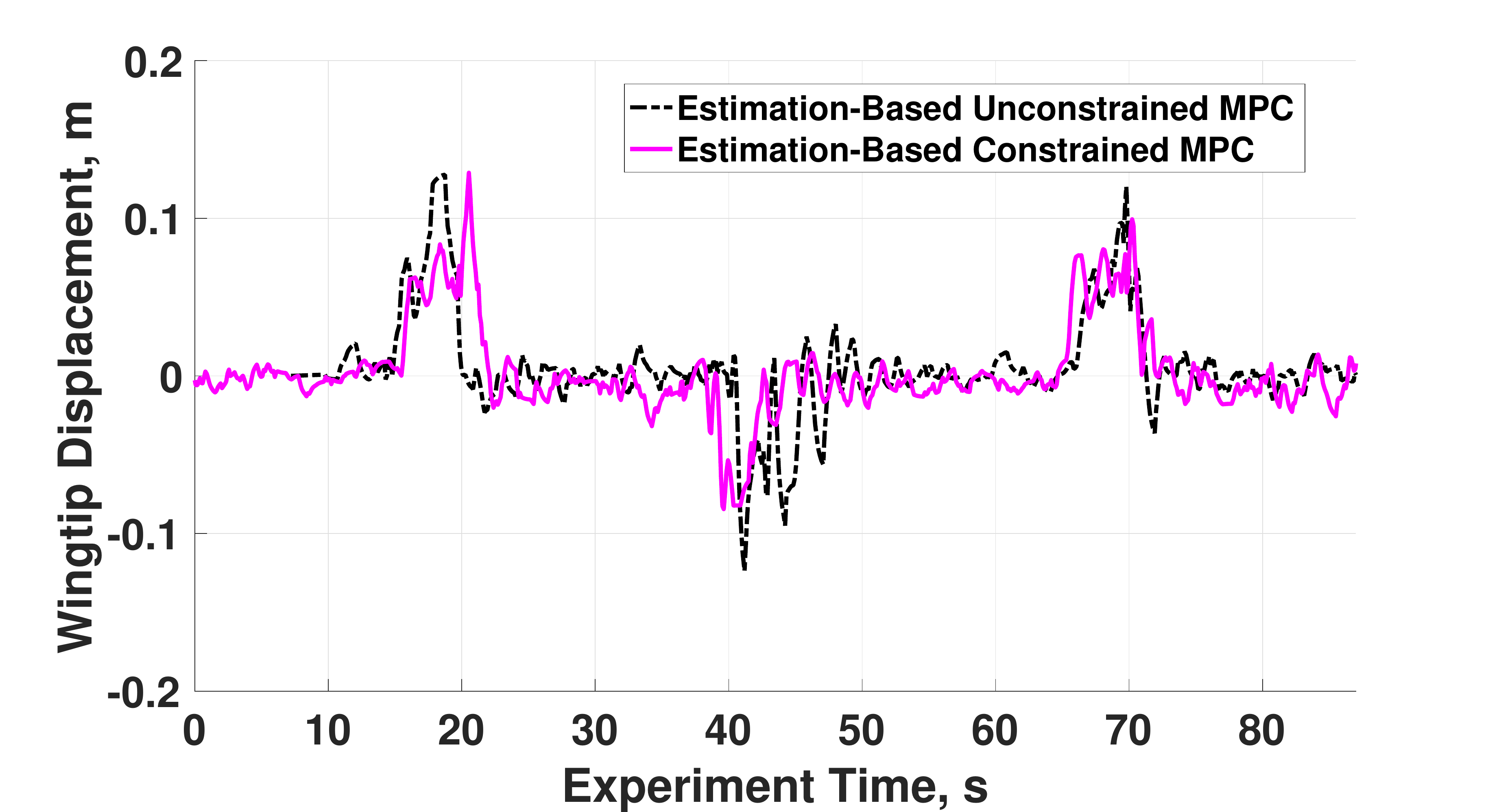} 
\end{center}
\caption{The wingtip displacement for a square disturbance resulting from putting on and off a weight of $15$~lb at the left wingtip every $25$\,s starting from time $t=15$\,s. The proposed estimation-based, unconstrained MPC strategy achieves a significantly faster response and much less peak and oscillations, compared to the basic PID controller, and almost the same behavior as the constrained MPC, while being much less computationally demanding.}
\label{constrainedMPC_versus_unconstrained_square}
\end{figure}

We next test the unconstrained MPC strategy, and compare it to the constrained MPC strategy. Figures \ref{constrainedMPC_versus_unconstrained} and \ref{constrainedMPC_versus_unconstrained_square} compare the estimation-based, constrained MPC response with the estimation-based, unconstrained MPC response for constant and square weight disturbances, respectively. The proposed estimation-based, unconstrained MPC strategy results in a significantly better performance than the PID controller in Figures \ref{fig_constant_disp}, \ref{fig_square_disp} and in a slightly worse performance than the estimation-based, constrained MPC strategy. On the other hand, the unconstrained MPC strategy does not require solving online optimization problems and, consequently, is less computationally demanding than the constrained MPC strategy. 

%
%
%
%

\subsection{Discussion of the Results}
\label{sec:exp_dis}
In this paper, we have presented several experimental results that show the effectiveness of our proposed estimation-based MPC strategy and the significant improvement it achieves over the basic PID control strategy. The results also verify the effectiveness of the different controller/estimator design approaches discussed in this paper and provide comparisons between these approaches. In particular, we have shown the following important points through our experimental results:
\begin{itemize}
\item Our proposed estimation-based MPC strategy speeds up the response of the basic PID controller around ten times, as shown in Figure \ref{fig_constant_disp}, without the need for adding any sensors. Hence, unlike the basic PID control strategy, the estimation-based MPC strategy enables the counter gust system to handle fast-changing disturbances (wind gusts) as shown in Figure \ref{fig_square_disp}. This faster response is expected since, in our approach, we estimate the wind torque and directly counteract its effect through the feed-forward compensation instead of waiting until the wind torque affects the roll angle to start counteracting its effect using the feedback controller. 
\item For the considered application, both the pole placement technique and the Kalman filter technique can be efficiently used to design the estimator gain, and both techniques achieve good estimation results as shown in Figure \ref{fig_wind_est_exp}.
\item For the considered application, the unconstrained MPC strategy achieves an excellent balance between optimality in performance and computational efficiency. In particular, it achieves lower peak values, less oscillations, and a significantly smaller settling time compared to the basic PID controller and a slightly worse performance than the constrained MPC strategy, as shown in Figures \ref{constrainedMPC_versus_unconstrained} and \ref{constrainedMPC_versus_unconstrained_square}. On the other hand, the unconstrained MPC strategy is much less computationally demanding than the constrained MPC strategy since it does not require solving online optimization problems.
\end{itemize}

For future research, we first plan to carry out extensive field tests, in which the vehicle is taxiing subject to crosswinds and wind gusts, instead of using weights attached to the wingtips for emulating wind torques. This will require buying more powerful motors that can counteract high wind speeds. Second, we are currently working on the control of a $3.5$-meter, fully autonomous version of the hybrid aerial vehicle. Third, for the current estimator design, we assume that the wind torque is constant during our short prediction horizon (around 3~s), and then counteract this estimated, constant wind torque using feed-forward compensation. It may be possible to create a wind prediction model that observes the past and current wind estimates from the Kalman filter as well as other environmental variables (e.g., ambient temperature and pressure), and then uses these values to predict the wind behavior over the prediction horizon. The predicted wind disturbance over the prediction horizon should be incorporated in the MPC for more accurate predictions and hence better overall response. Since the future wind predictions are typically probabilistic, stochastic MPC \cite{stochastic_mpc} or scenario-based MPC \cite{scenario_mpc} may be needed.

\section{Conclusions}
\label{sec:con}
In this paper, we designed and implemented an automatic crosswind  stabilization system for a novel hybrid aerial vehicle design, which enables take-off and landing in short distances without the need for runways. Despite its advantages, the vehicle's wing has a large surface area, which makes it very susceptible to crosswinds and wind gusts. 
The proposed automatic crosswind stabilization system detects the roll angle deviation, and then controls motors mounted at the vehicle's wingtips to counteract the wind effect. We have shown that standard control strategies such as a PID controller do not achieve fast stabilization of the wing and, consequently, cannot counteract wind gusts.
Given the difficulties of mounting reliable wind sensors on the considered wing, we used instead a wind torque estimator and provided two methods for the design of the estimator, pole placement and a Kalman filter. Then, we proposed an estimation-based predictive control strategy and showed through experimental results that the proposed strategy reduces the response time of the system by around $80$-$90\%$ compared to the PID controller without adding sensors or changing the hardware of the crosswind stabilization system. 

\bibliographystyle{IEEEtranS}

\end{document}